%% file: main.tex
\documentclass[journal]{IEEEtran}

\usepackage[T1]{fontenc}
\usepackage[utf8]{inputenc}
\usepackage{amsmath,amssymb,amsthm}
\usepackage{algorithm}
\usepackage{algorithmic}
\usepackage[most]{tcolorbox}
\usepackage{hyperref}
\usepackage{graphicx}
\usepackage[table]{xcolor}
\usepackage{booktabs}
\usepackage{newfloat}
\usepackage{caption}
\usepackage{url}
\usepackage{listings}
\usepackage{xcolor}
\definecolor{listingbg}{HTML}{F8F8F8}
\definecolor{listinggray}{gray}{0.5}
\definecolor{listingpurple}{RGB}{150,0,120}
\lstdefinestyle{xedbjson}{
  backgroundcolor=\color{listingbg},
  basicstyle=\ttfamily\footnotesize,
  numbers=left,
  numberstyle=\tiny\color{listinggray},
  numbersep=8pt,
  stepnumber=1,
  showstringspaces=false,
  frame=single,
  framexleftmargin=8pt,
  rulecolor=\color{listingbg},
  framerule=0pt,
  columns=fullflexible,
  keywordstyle=\color{listingpurple},
  breaklines=true,
  breakatwhitespace=true,
  literate=
    {\"}{{\color{listingpurple}{\"}}}1
    {:}{{\color{listinggray}{:}}}1
    {,}{{\color{listinggray}{,}}}1
    {[}{{\color{listinggray}{[}}}1
    {]}{{\color{listinggray}{]}}}1
    {\{}{{\color{listinggray}{\{}}}1
    {\}}{{\color{listinggray}{\}}}}1
}
\usepackage{xcolor} 
\usepackage{tikz}
\usetikzlibrary{arrows.meta,positioning,shapes.misc}

\definecolor{jsonbg}{HTML}{F8F9FB}
\definecolor{jsonframe}{HTML}{D8DEE9}
\definecolor{jsonstring}{HTML}{1F6F8B}
\definecolor{jsonbool}{HTML}{7A3E9D}
\definecolor{jsonpunct}{HTML}{5F6B7A}

\lstdefinestyle{xedbjson}{
  basicstyle=\ttfamily\footnotesize,
  backgroundcolor=\color{jsonbg},
  frame=single,
  rulecolor=\color{jsonframe},
  framerule=0.4pt,
  framesep=6pt,
  xleftmargin=0.5em,
  xrightmargin=0.5em,
  aboveskip=0.9em,
  belowskip=0.9em,
  captionpos=b,
  breaklines=true,
  breakatwhitespace=true,
  showstringspaces=false,
  columns=fullflexible,
  keepspaces=true,
  stringstyle=\color{jsonstring},
  literate=
   *{true}{{{\color{jsonbool}true}}}{4}
    {false}{{{\color{jsonbool}false}}}{5}
    {null}{{{\color{jsonbool}null}}}{4}
    {:}{{{\color{jsonpunct}{:}}}}{1}
    {,}{{{\color{jsonpunct}{,}}}}{1}
}

\usepackage{caption}

\definecolor{codebg}{HTML}{FAFAFA}
\definecolor{codeframe}{HTML}{D9DEE7}
\definecolor{codekey}{HTML}{2F5D8C}
\definecolor{codeval}{HTML}{444444}
\definecolor{codebool}{HTML}{7A3E9D}

\lstdefinestyle{xedbyaml}{
  basicstyle=\ttfamily\scriptsize,
  backgroundcolor=\color{codebg},
  frame=single,
  rulecolor=\color{codeframe},
  framerule=0.4pt,
  framesep=5pt,
  xleftmargin=0.25em,
  xrightmargin=0.25em,
  aboveskip=0.75em,
  belowskip=0.75em,
  captionpos=b,
  breaklines=true,
  breakatwhitespace=true,
  showstringspaces=false,
  columns=fullflexible,
  keepspaces=true,
  literate=
    {true}{{{\color{codebool}true}}}{4}
    {false}{{{\color{codebool}false}}}{5}
    {:}{{{\color{codekey}:}}}{1}
}

\title{Testing Agentic Workflows with Structural Coverage Criteria}

\author{Nafiseh Kahani and Mojtaba Bagherzadeh
\thanks{First Author is with the Department of Systems and Computer Engineering, Carleton University, Ottawa, Canada. Email: nafisehkahani@cunet.carleton.ca}%
\thanks{Second Author is with the  Cisco Systems, Ottawa, Canada. Email: mbagherz@cisco.com}%
}

\DeclareFloatingEnvironment[
    fileext=lol,
    listname=List of Listings,
    name=Listing,
]{listing}

\begin{document}

\maketitle

\begin{abstract}
Multi-agent systems increasingly expose explicit workflow structure: agents, tools, tool-access rules, restrictions, and delegation paths. Existing evaluations rely largely on end-to-end task success, benchmark scores, final-response quality, or prompt-level checks, which provide limited evidence that this declared coordination structure has actually been exercised. This makes it difficult to assess test-suite adequacy or detect structural regressions in tool access, restrictions, and inter-agent delegation. We address this gap with a structural testing approach for multi-agent workflow specifications. The approach represents each workflow as a typed coordination graph, derives coverage obligations over reachable agents, allowed tool edges, restricted tool edges, and delegation edges, and uses coverage-driven generation with DSPy-based scenario realization to produce executable tests. The graph fixes what must be covered; DSPy realizes those obligations as natural-language scenarios whose witnesses are checked at runtime. We implement the approach for OpenAI Agents SDK-style workflows and evaluate it on ten SDK-derived benchmarks comprising 49 reachable agents, 47 tools, and 403 structural obligations. Generated scenarios witness 54/75 allowed-tool obligations and 36/48 delegation obligations within a bounded refinement budget. The adversarial restricted-tool criterion elicits 23/248 restricted-call violations, separating workflows whose restrictions hold under probing from workflows with concrete misrouting failures. These results show that structural coverage provides a useful adequacy layer for multi-agent workflow testing: it does not replace semantic or end-to-end evaluation, but reveals whether declared agents, tool-access rules, restrictions, and delegation paths have been exercised.

\end{abstract}


\input{introduction}
\input{background}
\input{approach}
\input{evaluation}
\input{related-work}
\input{discussion-conclusion}

\bibliographystyle{plain}
\bibliography{references}

\end{document}

%% file: introduction.tex
\section{Introduction}
\label{sec:introduction}

LLM-based agents are software systems that use language models to interpret context, reason about goals, and take actions on behalf of users, often through tools or external APIs \cite{xi2023rise}. In this paper, we focus on LLM-based software agents and refer to them simply as agents. Recent tool-augmented and multi-agent systems extend beyond single-turn prompting to support information retrieval, planning, API use, orchestration, and delegation across specialized agents \cite{yao2023react, wu2024autogen}. As these systems become more capable, they are increasingly deployed as structured workflows rather than isolated prompts: agents are assigned roles, connected to tools, constrained by tool-access rules, and linked through handoff or delegation paths.

Existing evaluation practice largely emphasizes observable task outcomes, benchmark performance, final response quality, and tool-use behavior \cite{liu2023agentbench, mialon2023gaia, li2023apibank, qin2023toolllm, jimenez2024swebench, zhou2024webarena}. These metrics are useful, but they provide limited evidence that a system's internal coordination structure has been covered. A multi-agent workflow may complete a task successfully even if some tools are never invoked, some restricted tool accesses are never tested, or some delegation paths remain unexplored. Thus, end-to-end task success is not a reliable proxy for structural coverage in multi-agent systems.

This gap becomes more important as multi-agent systems evolve. A small change to a workflow definition may add a new tool edge, remove a restriction, or alter a delegation path without causing immediate end-to-end failures. Conventional evaluations may still pass while important structural obligations remain unobserved. This makes it difficult to assess whether a test suite is adequate, whether a system change has been meaningfully exercised, and whether regressions have been introduced into the workflow's coordination structure. The issue is especially important in safety-, policy-, or compliance-sensitive settings, where practitioners often need traceable evidence that relevant tool-access rules, restrictions, and coordination paths have been tested.

We therefore treat multi-agent testing as both a behavioral and structural problem. We model a multi-agent workflow as a typed coordination graph whose nodes represent agents and tools, and whose edges represent relations such as allowed tool access, restricted tool access, and delegation. This perspective follows established ideas in software testing, where adequacy criteria such as statement coverage, branch coverage, and mutation-based testing are used to assess how thoroughly a program has been exercised \cite{ammann2016introduction, myers2011art}. Related work on behavioral testing has likewise shown that aggregate accuracy or task success can mask important coverage gaps \cite{ribeiro2020checklist, wang2021textflint}. However, structural adequacy for multi-agent workflows remains comparatively underdeveloped.

In this paper, we define a structural adequacy model for typed multi-agent coordination graphs. The model includes four coverage criteria spanning reachable agents, allowed tool edges, restricted tool edges, and delegation edges. These criteria measure not only whether a multi-agent system produces acceptable outputs, but also whether its declared coordination structure has been witnessed during testing. We then introduce a coverage-driven generation procedure that synthesizes structural witness objectives from the graph and uses DSPy modules to realize each objective as a natural-language test scenario \cite{khattab2023dspy,dspy_docs}. The structural meaning of each test is fixed by the graph; DSPy is used only for controlled scenario realization, and runtime witnesses determine whether a generated scenario actually covers its intended obligation.

We implement the approach as a prototype for OpenAI Agents SDK-style workflows. The pipeline takes a Python agent entry point, extracts a normalized workflow manifest, derives graph-based structural obligations, realizes typed witness objectives as natural-language scenarios, and measures coverage by replaying those scenarios against a runtime adapter. The implementation supports both one-shot execution and stage-by-stage invocation, so extraction, obligation construction, scenario realization, and runtime evaluation can be inspected independently.

We evaluate the approach on ten SDK-derived workflows extracted from the OpenAI Agents SDK examples and public third-party projects built on the same SDK. Together, these benchmarks contain 49 reachable agents, 47 tools, and 403 graph-derived structural obligations. The evaluation is organized around four research questions: whether the pipeline can extract valid typed coordination graphs and structural obligations from SDK-derived workflows (RQ1); how often generated scenarios witness allowed-tool and delegation obligations at runtime (RQ2); whether adversarial restricted-tool testing reveals concrete violations of declared restrictions (RQ3); and whether the same structural harness can test robustness under tool faults on a instrumented benchmark (RQ4).

Our results show that structural obligations can be extracted and exercised across workflows of varying size and coordination structure. Across the runtime evaluation, generated scenarios witness 54/75 allowed-tool obligations and 36/48 delegation obligations within a bounded refinement budget. The adversarial restricted-tool criterion elicits 23/248 restricted-call violations, separating workflows whose restrictions hold under probing from workflows with concrete misrouting failures. We use the OpenAI Agents SDK \texttt{oai\_customer\_service} workflow as both a running example throughout the paper and a instrumented benchmark for routing-prompt ablation and tool-fault robustness. These results show that structural testing complements existing agent evaluations: it does not replace end-to-end benchmarks or semantic quality assessment, but provides explicit evidence that a multi-agent workflow's declared agents, tool-access rules, restrictions, and delegation paths have been exercised.

The paper makes the following contributions:
\begin{itemize}
\item We formulate testing of multi-agent workflow specifications as a structural adequacy problem over typed coordination graphs derived from agent-framework source code.
\item We define graph-derived coverage criteria for reachable agents, allowed tool edges, restricted tool edges, and delegation edges. Restricted-tool coverage is treated as an explicit negative-observation obligation rather than as the mere absence of a tool call.
\item We present a coverage-driven test-generation pipeline that converts workflow manifests into typed witness objectives and uses DSPy-based, runtime-grounded realization to produce natural-language test scenarios.
\item We evaluate the approach on ten OpenAI Agents SDK-derived workflows totaling 403 structural obligations, including a deeper analysis of \texttt{oai\_customer\_service} for routing-prompt ablation and tool-fault robustness.
\end{itemize}

The remainder of the paper is organized as follows. Section~II defines multi-agent workflow specifications and introduces the \texttt{oai\_customer\_service} running example. Section~III presents the typed coordination-graph model, the four structural coverage criteria, and the coverage-driven generation procedure. Section~IV describes the benchmark suite, implementation, and empirical evaluation across ten SDK-derived workflows. Section~V discusses related work in structural testing, model-based testing, automated test generation, and LLM-agent evaluation. Section~VI discusses limitations, future work, and concludes.

%% file: background.tex
\section{Multi-Agent Workflow Specifications}
\label{sec:background}

In this work, a \emph{multi-agent workflow specification} is a normalized description of the structural elements of an LLM-based multi-agent workflow that are relevant for graph-based testing. It records the agents that may participate in an execution, the tools associated with the workflow, the tool-access rules that determine which agents may or may not invoke which tools, and the delegation paths by which control may move between agents. We use this term to refer to the representation analyzed by our testing method, rather than to a universal format shared by all agent frameworks.

The core declarations are:
\begin{itemize}
    \item a set of \textbf{reachable agents}, meaning agents that can be reached from the workflow entry point through zero or more delegation steps;
    \item \textbf{allowed tool edges}, which record which tools each agent may invoke;
    \item \textbf{restricted tool edges}, which record tools that an agent is explicitly restricted from invoking; and
    \item \textbf{delegation edges}, which describe how one agent may hand control or responsibility to another agent.
\end{itemize}

We interpret tool access using a closed-world convention. A tool is available to an agent only when an explicit allowed-tool edge is declared. A restricted-tool edge represents an explicit negative obligation: generated tests should witness that the workflow respects the restriction. If no edge is declared between an agent and a tool, the pair is treated as unspecified rather than both allowed and restricted. Such pairs do not induce coverage obligations in our model. We also require well-formed specifications to avoid contradictory declarations, so the same agent--tool pair cannot appear as both allowed and restricted.

We deliberately keep execution-time concerns out of the specification graph. Whether a particular tool is implemented as a Python function, an MCP server, or an HTTP endpoint, and whether the harness runs the test in a validation, dry-run, or full-execution mode, are properties of the runtime layer. They affect how a test is observed but do not change the set of structural obligations declared by the specification. We therefore treat the specification graph as a static description of agents, tools, allowed tool edges, restricted tool edges, and delegations.

\subsection{Running Example: \texttt{oai\_customer\_service}}
\label{sec:running-example}

We use \texttt{oai\_customer\_service} as the main running example for introducing the workflow specification and graph-derived coverage model. The workflow is extracted from the public OpenAI Agents SDK repository~\cite{openai_agents_sdk}; its source is the \texttt{examples/customer\_service/main.py} file of \texttt{openai/openai-agents-python}. We extract the manifest mechanically, by importing the example module and introspecting its \texttt{Agent}, \texttt{function\_tool}, and \texttt{handoff} objects. The specification used here is therefore a faithful normalization of what an SDK developer would write, not a hand-crafted toy. We later use \texttt{oai\_message\_filter} in Section~\ref{sec:dspy-realization} only to illustrate a multi-attempt delegation-refinement trace, because that benchmark exposes the runtime-feedback behavior of the realizer more clearly than the simpler customer-service example.

The workflow contains three reachable agents:
\begin{itemize}
    \item \texttt{triage\_agent} is the entry-point agent. It owns no tools of its own and acts as a router that hands the conversation off to the agent best suited to the user's request.
    \item \texttt{faq\_agent} answers frequently asked questions about flights and is permitted to call the \texttt{faq\_lookup\_tool} function tool.
    \item \texttt{seat\_booking\_agent} updates a passenger's seat assignment and is permitted to call the \texttt{update\_seat} function tool.
\end{itemize}

There are two function tools, \texttt{faq\_lookup\_tool} and \texttt{update\_seat}. Each tool is allowed for exactly one agent and explicitly restricted for the other two. In addition, both leaf agents are configured to hand the conversation back to \texttt{triage\_agent} once their specialized work is done, so the delegation structure is bidirectional rather than tree-shaped.

Figure~\ref{fig:running-example-graph} shows the corresponding coordination graph. Rounded nodes denote agents, rectangular nodes denote tools, solid edges denote allowed tool access, and dotted edges denote delegation. For readability, the figure shows only the two allowed-tool edges and the four delegation edges; the four restricted-tool edges are listed in the caption.

\begin{figure}[tbp]
\centering
\begin{tikzpicture}[
  >=Latex,
  agent/.style={draw, rounded corners, thick, minimum width=2.0cm, minimum height=0.8cm, align=center, font=\small},
  tool/.style={draw, rectangle, thick, minimum width=2.2cm, minimum height=0.8cm, align=center, font=\small},
  allow/.style={->, thick},
  delegate/.style={->, thick, dotted}
]

\node[agent] (triage) {triage\_agent};
\node[agent, below left=1.6cm and 0.2cm of triage] (faq) {faq\_agent};
\node[agent, below right=1.6cm and 0.2cm of triage] (seat) {seat\_booking\_agent};

\node[tool, below=1.0cm of faq] (faqtool) {faq\_lookup\_tool};
\node[tool, below=1.0cm of seat] (seattool) {update\_seat};

\draw[delegate, transform canvas={xshift=-2pt}] (triage) -- node[left, font=\scriptsize] {delegate} (faq);
\draw[delegate, transform canvas={xshift=2pt}]  (faq) -- (triage);
\draw[delegate, transform canvas={xshift=2pt}]  (triage) -- node[right, font=\scriptsize] {delegate} (seat);
\draw[delegate, transform canvas={xshift=-2pt}] (seat) -- (triage);

\draw[allow] (faq) -- node[left, font=\scriptsize] {allow} (faqtool);
\draw[allow] (seat) -- node[right, font=\scriptsize] {allow} (seattool);

\end{tikzpicture}
\caption{Coordination graph for the \texttt{oai\_customer\_service} running example. Rounded nodes are agents, rectangular nodes are tools, solid edges denote allowed tool access, and dotted edges denote delegation. The four restricted-tool edges, omitted from the figure for clarity, are: \texttt{(triage\_agent, faq\_lookup\_tool)}, \texttt{(triage\_agent, update\_seat)}, \texttt{(faq\_agent, update\_seat)}, and \texttt{(seat\_booking\_agent, faq\_lookup\_tool)}.}
\label{fig:running-example-graph}
\end{figure}

Listing~\ref{lst:running-example-spec} shows the normalized specification used by the prototype. The listing uses a YAML rendering for readability.

\begin{lstlisting}[
  style=xedbyaml,
  float=tbp,
  caption={Normalized specification for the \texttt{oai\_customer\_service} running example, extracted from the OpenAI Agents SDK \texttt{examples/customer\_service} module.},
  label={lst:running-example-spec}
]
system:
  id: oai_customer_service
  entry_agent: triage_agent

agents:
  - id: triage_agent
  - id: faq_agent
  - id: seat_booking_agent

tools:
  - id: faq_lookup_tool
  - id: update_seat

permissions:
  allow:
    - [faq_agent,          faq_lookup_tool]
    - [seat_booking_agent, update_seat]

  restrict:
    - [triage_agent,       faq_lookup_tool]
    - [triage_agent,       update_seat]
    - [faq_agent,          update_seat]
    - [seat_booking_agent, faq_lookup_tool]

delegations:
  - {from: triage_agent,       to: faq_agent,          trigger: delegate}
  - {from: faq_agent,          to: triage_agent,       trigger: delegate}
  - {from: triage_agent,       to: seat_booking_agent, trigger: delegate}
  - {from: seat_booking_agent, to: triage_agent,       trigger: delegate}
\end{lstlisting}

We return to this example throughout the paper. Section~\ref{sec:formal-coverage} uses it to illustrate the formal coverage model, Section~\ref{sec:generate} traces the generation procedure on it, and Section~\ref{sec:evaluation} uses it as the instrumented benchmark within a broader ten-workflow evaluation.

\subsection{DSPy for Test Scenario Realization}
\label{sec:dspy-background}

Our structural coverage model is symbolic: obligations are derived from a typed coordination graph. However, executing those obligations against LLM-based agent systems still requires natural-language scenarios. To generate those scenarios in a controlled way, we use DSPy as a scenario-realization layer rather than as the source of the coverage model.

DSPy is a declarative framework for programming language-model pipelines using typed signatures, modules, and optimization metrics rather than hand-written prompt strings \cite{khattab2023dspy,dspy_docs}. A DSPy signature specifies the input--output behavior of a language-model module, while a module implements a reusable language model call that can be composed with other modules \cite{dspy_signatures,dspy_modules}. DSPy programs can also be optimized against user-defined metrics, which makes them useful when the desired output can be checked automatically or semi-automatically \cite{khattab2023dspy}.

In our setting, DSPy is used only after the structural obligations have been computed. The graph model determines what must be covered; DSPy helps realize each obligation as a concrete test scenario. This separation is important: the adequacy criteria do not depend on the language model's judgment. Instead, the language model proposes candidate scenarios, and deterministic checks verify that each scenario is aligned with the intended obligation, consistent with the workflow specification, and paired with observable structural evidence.

%% file: approach.tex
\newcommand{\TRC}[1]{#1}
\section{Approach}
\label{sec:approach}

We propose a \emph{structural testing} pipeline for multi-agent workflows. Rather than judging a workflow only by end-to-end task success, the pipeline first reads the workflow specification as a structured graph, extracts the structural obligations declared by that graph, and generates tests that explicitly try to witness those obligations. In our setting, the obligations concern reachable agents, allowed tool edges, restricted tool edges, and delegation edges.

The approach follows the five-step pipeline shown in Figure~\ref{fig:approach-overview}. \textcircled{1} We begin from a normalized multi-agent workflow specification $S$. \textcircled{2} We convert $S$ into a reachable typed coordination graph $G(S)$. \textcircled{3} From $G(S)$, we derive graph-based obligations. \textcircled{4} We synthesize typed witness objectives and use DSPy to realize them as natural-language test scenarios. \textcircled{5} We execute the resulting tests through a runtime adapter and measure whether the expected structural evidence is observed. Overall, the method follows a \emph{specification $\rightarrow$ graph $\rightarrow$ obligations $\rightarrow$ tests $\rightarrow$ observations} pipeline.

The running example introduced in Section~\ref{sec:running-example} provides a concrete instance of this process. It contains three reachable agents, two tools, two allowed tool edges, four restricted tool edges, and four delegation edges, yielding 13 structural obligations. Section~\ref{sec:evaluation} later evaluates the same procedure across ten SDK-derived workflows.

\begin{figure}[t]
\centering
\caption{Overview of the proposed pipeline. A normalized multi-agent workflow specification $S$ is converted into a reachable coordination graph $G(S)$, from which structural obligations are extracted. These obligations are turned into coverage-driven witness objectives and realized as executable tests via DSPy. Runtime observations then support coverage measurement.}
\label{fig:approach-overview}
\small
\setlength{\tabcolsep}{6pt}
\begin{tabular}{c}
\fbox{\parbox{0.9\columnwidth}{\centering
\textbf{\textcircled{1} Normalized Workflow Specification $S$}\\
Agents, tools, allow/restrict edges, delegation edges}}\\[0.6ex]
$\downarrow$\\[0.6ex]
\fbox{\parbox{0.9\columnwidth}{\centering
\textbf{\textcircled{2} Reachable Coordination Graph $G(S)$}\\
Typed nodes and reachable structural relations}}\\[0.6ex]
$\downarrow$\\[0.6ex]
\fbox{\parbox{0.9\columnwidth}{\centering
\textbf{\textcircled{3} Structural Obligations}\\
Reachable agents, allowed tools, restricted tools, delegations}}\\[0.6ex]
$\downarrow$\\[0.6ex]
\fbox{\parbox{0.9\columnwidth}{\centering
\textbf{\textcircled{4} Generated Structural Tests}\\
Natural-language scenarios with expected structural witnesses}}\\[0.6ex]
$\downarrow$\\[0.6ex]
\fbox{\parbox{0.9\columnwidth}{\centering
\textbf{\textcircled{5} Execution and Coverage Measurement}\\
Runtime observations, witness verdicts, coverage scores}}
\end{tabular}
\end{figure}

\subsection{Formal Coverage Model}
\label{sec:formal-coverage}

Let $S$ be a normalized multi-agent workflow specification. We model the structurally relevant content of $S$ as a typed coordination graph
\[
G(S) =
\bigl(
V_A,\,
V_T,\,
E_{\mathit{allow}},\,
E_{\mathit{restrict}},\,
E_{\mathit{del}}
\bigr),
\]
where $V_A$ is the set of agent nodes, $V_T$ is the set of tool nodes, $E_{\mathit{allow}} \subseteq V_A \times V_T$ is the set of allowed agent--tool edges, $E_{\mathit{restrict}} \subseteq V_A \times V_T$ is the set of restricted agent--tool edges, and $E_{\mathit{del}} \subseteq V_A \times V_A$ is the set of delegation edges.

Tools are modeled as opaque nodes. The graph does not record how a tool is implemented, such as a Python function, an MCP server, or an HTTP endpoint, nor which execution mode the harness uses to invoke it. These are runtime concerns that affect \emph{how} a witness is observed but not \emph{which} structural obligations exist.

Let $A_{\mathit{reach}} \subseteq V_A$ be the set of agents reachable from the workflow entry point by following zero or more delegation edges. Reachability is computed over $E_{\mathit{del}}$, so cycles in the delegation graph do not pose a problem. Structural coverage is defined only over reachable parts of the workflow. We assume well-formed specifications: all edge endpoints must be declared, and no agent--tool pair may appear in both $E_{\mathit{allow}}$ and $E_{\mathit{restrict}}$.

From $G(S)$ we derive three additional reachable obligation sets:
\begin{align*}
T_{\mathit{allow}} &=
\{(a,t) \in E_{\mathit{allow}} \mid a \in A_{\mathit{reach}}\}, \\
T_{\mathit{restrict}} &=
\{(a,t) \in E_{\mathit{restrict}} \mid a \in A_{\mathit{reach}}\}, \\
D &=
\{(a,b) \in E_{\mathit{del}} \mid a \in A_{\mathit{reach}} \land b \in A_{\mathit{reach}}\}.
\end{align*}

Thus, the structural obligation space is
\[
\Omega(S) =
\bigl(
A_{\mathit{reach}},\,
T_{\mathit{allow}},\,
T_{\mathit{restrict}},\,
D
\bigr).
\]

Each executed test $x$ produces a structural observation
\[
\mathsf{obs}(x) =
\bigl(
\mathit{agents}(x),\,
\mathit{tools}(x),\,
\mathit{restricted}(x),\,
\mathit{delegations}(x)
\bigr),
\]
where $\mathit{agents}(x)$ records observed agents, $\mathit{tools}(x)$ records observed allowed tool invocations, $\mathit{restricted}(x)$ records explicit restricted-access outcomes, and $\mathit{delegations}(x)$ records observed delegation edges.

A restricted-tool obligation is not witnessed merely by the absence of a tool call. It requires an explicit restricted-access observation, such as a harness-level rejection, policy event, refusal trace, or attempted restricted call recorded by the runtime adapter. This distinction is important because absence alone does not show that the restriction was actually exercised.

For a finite obligation set $O$ and a witness predicate $W(x,\omega)$, we define coverage as
\[
\mathsf{Cov}(X,O,W) =
\begin{cases}
1, & \text{if } O = \varnothing, \\[0.75ex]
\dfrac{\bigl|\{\omega \in O \mid \exists x \in X : W(x,\omega)\}\bigr|}{|O|}, & \text{otherwise,}
\end{cases}
\]
where $X$ is the generated test suite. This convention treats criteria with no applicable obligations as vacuously satisfied.

\paragraph{C1. Agent coverage}
Every reachable agent is observed by at least one test:
\[
C_1(X,S) =
\mathsf{Cov}\bigl(
X,\,
A_{\mathit{reach}},\,
\lambda(x,a).\, a \in \mathit{agents}(x)
\bigr).
\]

\paragraph{C2. Allowed-tool coverage}
Every reachable allowed agent--tool edge is exercised:
\[
C_2(X,S) =
\mathsf{Cov}\bigl(
X,\,
T_{\mathit{allow}},\,
\lambda(x,(a,t)).\, (a,t) \in \mathit{tools}(x)
\bigr).
\]

\paragraph{C3. Restricted-tool coverage}
Every reachable restricted agent--tool edge is exercised through an explicit restricted-access observation:
\[
\begin{aligned}
C_3(X,S)
&= \mathsf{Cov}\bigl(
    X,\,
    T_{\mathit{restrict}}, \\
&\qquad
    \lambda(x,(a,t)).\, (a,t) \in \mathit{restricted}(x)
\bigr).
\end{aligned}
\]
For example, a request that tempts \texttt{triage\_agent} to update a seat directly should produce explicit evidence that direct use of \texttt{update\_seat} by \texttt{triage\_agent} is blocked, refused, redirected, or otherwise recorded as a restricted-access event.

\paragraph{C4. Delegation coverage}
Every reachable delegation edge is observed:
\[
C_4(X,S) =
\mathsf{Cov}\bigl(
X,\,
D,\,
\lambda(x,(a,b)).\, (a,b) \in \mathit{delegations}(x)
\bigr).
\]

Full structural coverage means that all four criteria are satisfied. It does not, by itself, guarantee semantic correctness, safety, or output quality; it only shows that the declared coordination structure has been exercised.

\subsection{Coverage-Driven Test Generation}
\label{sec:generate}

We now describe how tests are generated from the graph-derived obligation space. The guiding principle is one obligation, at least one witness objective. Rather than synthesizing arbitrary task prompts and hoping they exercise the relevant parts of a workflow, the generator starts from the obligation sets and constructs focused objectives designed to witness them directly.

The generator has two layers. The first layer is deterministic: it constructs the reachable coordination graph, extracts obligations, and converts each obligation into a typed witness objective. The second layer is DSPy-based: it realizes each witness objective as a natural-language test scenario, checks whether the scenario is aligned with the intended obligation, repairs invalid scenarios when possible, and emits executable tests with expected structural observables. Thus, DSPy is used for controlled scenario realization, while the coverage model remains graph-defined \cite{khattab2023dspy,dspy_docs}.

Formally, the generator is a mapping
\[
\mathsf{Generate}(S) \rightarrow X,
\]
where $S$ is a normalized multi-agent workflow specification and $X$ is a structural test suite. Generation proceeds in five phases:
\begin{enumerate}
    \item \textbf{Graph construction.} Parse $S$ and compute $G(S)$.
    \item \textbf{Obligation extraction.} Derive $\Omega(S) = (A_{\mathit{reach}}, T_{\mathit{allow}}, T_{\mathit{restrict}}, D)$.
    \item \textbf{Objective construction.} Convert each obligation into a typed witness objective.
    \item \textbf{Scenario realization.} Use DSPy modules to realize objectives as natural-language scenarios.
    \item \textbf{Validation and assembly.} Check candidates against structural constraints and emit executable tests.
\end{enumerate}

Each generated test specification contains a target agent or workflow entry point, one or more structural witness objectives, a natural-language user scenario, and the expected structural evidence: observed agents, agent--tool uses, restricted-access outcomes, and delegation edges.

\paragraph{Witness objectives}
A witness objective is a typed request for evidence about one structural obligation. We use four objective families:
\[
\mathsf{Reach}(a),\;
\mathsf{UseTool}(a,t),\;
\mathsf{RestrictTool}(a,t),\;
\mathsf{Delegate}(a,b).
\]
They ask, respectively, for evidence that agent $a$ is reached, that agent $a$ uses allowed tool $t$, that restricted access from $a$ to $t$ is explicitly observed, and that control transfers from agent $a$ to agent $b$.

\paragraph{Objective construction rules}
The deterministic layer constructs objectives using four rules:
\begin{itemize}
    \item \textbf{R1 (Agent rule).} For each $a \in A_{\mathit{reach}}$, emit $\mathsf{Reach}(a)$.
    \item \textbf{R2 (Allowed-tool rule).} For each $(a,t) \in T_{\mathit{allow}}$, emit $\mathsf{UseTool}(a,t)$.
    \item \textbf{R3 (Restricted-tool rule).} For each $(a,t) \in T_{\mathit{restrict}}$, emit $\mathsf{RestrictTool}(a,t)$.
    \item \textbf{R4 (Delegation rule).} For each $(a,b) \in D$, emit $\mathsf{Delegate}(a,b)$.
\end{itemize}
Applied to \texttt{oai\_customer\_service}, these rules emit $3 + 2 + 4 + 4 = 13$ typed witness objectives.

\paragraph{Merging.}
A single test may witness multiple obligations. For example, a test that exercises an allowed tool of agent $a$ may also witness that $a$ was reached. The merge step therefore combines compatible objectives into \emph{objective bundles} to reduce suite size. Each bundle contains one \emph{driver objective}, which determines the DSPy realization signature, and zero or more secondary objectives, which are checked as additional witness obligations on the same runtime trace.

A merge is allowed only when it is \emph{obligation-preserving}: the realized test must retain the witness obligations of every objective in the bundle. In the current implementation, heterogeneous bundles are allowed only when the secondary objectives are reachability objectives entailed by the driver. For example, a bundle containing $\mathsf{Reach}(a)$ and $\mathsf{UseTool}(a,t)$ is realized using the $\mathsf{RealizeUseTool}$ signature, and the resulting trace is checked for both reaching $a$ and observing the tool-use edge $(a,t)$. We do not merge objectives with different non-reachability drivers, such as $\mathsf{UseTool}(a,t)$ and $\mathsf{Delegate}(a,b)$, because they require different realization constraints. We also forbid merges that introduce contradictions, such as requiring the same agent to both use and be restricted from using the same tool in one execution.

\subsection{DSPy-Based Scenario Realization}
\label{sec:dspy-realization}

Each typed witness objective is realized by a DSPy module that produces a candidate end-user message and pairs it with the structural evidence expected from executing it. The structural meaning of a test is fixed by the objective; DSPy is responsible only for surface form and contextual plausibility.

\paragraph{One signature per criterion}
The realizer is not a single free-form prompt template. Each coverage criterion has its own DSPy signature \cite{dspy_signatures}, with structured input fields lifted from the manifest and criterion-specific generation constraints:
\begin{itemize}
    \item $\mathsf{RealizeReach}(C_1)$ produces a realistic request that should route to the target agent without naming it.
    \item $\mathsf{RealizeUseTool}(C_2)$ produces a request that requires the capability provided by the target tool without naming the tool.
    \item $\mathsf{RealizeRestrictTool}(C_3)$ produces a request that tempts an agent toward a restricted capability while staying within its surface domain.
    \item $\mathsf{RealizeDelegate}(C_4)$ produces a request that should arrive at a parent agent but require delegation to a child agent.
\end{itemize}

This separation matters because the objectives have different and sometimes opposite goals. For example, $C_2$ asks the realizer to elicit a permitted tool call, while $C_3$ asks it to probe a forbidden boundary. A single generic prompt template would not express these distinctions cleanly.

\paragraph{Bounded refinement with an execution-grounded reward}
Surface plausibility alone is a poor proxy for structural coverage: a well-written user message can be answered by the wrong agent or without the intended tool call. We therefore wrap the per-criterion realizer in a bounded \texttt{dspy.Refine} loop \cite{dspy_docs}. Each candidate prompt $z$ is executed through the runtime adapter and scored by a binary reward:
\[
r(z) =
\begin{cases}
0, &
\begin{aligned}[t]
&\text{if } z \text{ leaks a forbidden identifier,}\\
&\text{such as a literal tool or agent name,}
\end{aligned}
\\[1.0ex]
1, &
\begin{aligned}[t]
&\text{if executing } z \text{ produces the structural}\\
&\text{evidence required by the objective,}
\end{aligned}
\\[1.0ex]
0, & \text{otherwise.}
\end{cases}
\]

The loop terminates as soon as $r(z)=1$, or after $N$ attempts. In our experiments, $N=5$. After each attempt, the witness verdict and observed trace are fed back to the realizer, allowing it to revise based on runtime behavior rather than its own confidence. If no candidate succeeds within the budget, the objective is marked unrealized and reported.

\textbf{Worked example: a delegation realization.}
Figure~\ref{fig:dspy-trace} shows a real $\mathsf{RealizeDelegate}$ run on the \texttt{oai\_message\_filter} workflow, for the obligation $\mathsf{Delegate}(\texttt{assistant\_2}\to\texttt{spanish\_assistant})$. Although \texttt{oai\_customer\_service} remains the paper's main running example, we use \texttt{oai\_message\_filter} here because it provides a clearer illustration of multi-attempt runtime-grounded refinement. Early attempts ask for Spanish output but are answered directly by \texttt{assistant\_2}, so no handoff is observed. After runtime feedback, the realizer changes the prompt to explicitly allow connection to someone who can assist in Spanish; this produces the intended handoff.

\begin{figure}[t]
\small
\setlength{\fboxsep}{6pt}
\framebox[\linewidth]{%
\parbox{0.95\linewidth}{%
\textbf{Objective.} $\mathsf{Delegate}(\texttt{assistant\_2}\to\texttt{spanish\_assistant})$ on \texttt{oai\_message\_filter}.\\
\textbf{Signature inputs.} Parent description of \texttt{assistant\_2}; child description of \texttt{spanish\_assistant}; delegation trigger ``user prefers Spanish.''\\[3pt]
\textbf{Attempt 1.} \emph{``¿Puedes ayudarme a traducir este párrafo al español, por favor?''}\quad runtime: \texttt{assistant\_2} replies in Spanish; no handoff; $r=0$.\\[2pt]
\textbf{Attempt 2.} \emph{``Hi, could you briefly explain this phrase in Spanish? \dots if my Spanish isn't perfect, please kindly pass me on to someone who can assist fluently in Spanish.''}\quad runtime: \texttt{assistant\_2} answers in English; no handoff; $r=0$.\\[2pt]
\textbf{Attempt 3.} \emph{``Please provide a very brief explanation of this text in Spanish \dots I'm more comfortable using that language for this.''}\quad runtime: \texttt{assistant\_2} replies in Spanish; no handoff; $r=0$.\\[2pt]
\textbf{Attempt 4.} \emph{``Could you give me a very brief summary of this text? I'm more comfortable with Spanish, so please reply in Spanish \textbf{or connect me with someone who can assist in Spanish}.''}\quad runtime: handoff \texttt{assistant\_2} $\to$ \texttt{spanish\_assistant} observed; $r=1$. Stop.
}}
\caption{A four-attempt $\mathsf{RealizeDelegate}$ trace from \texttt{oai\_message\_filter}. The realizer converges on a phrasing that produces the intended delegation only after runtime-grounded feedback.}
\label{fig:dspy-trace}
\end{figure}

The realizer cannot satisfy an objective merely by writing a plausible scenario. A candidate is accepted only when the runtime trace contains the expected structural evidence. This is what makes the generated suite coverage-driven rather than prompt-driven: the graph defines the obligation, DSPy proposes a scenario, and the runtime witness decides whether the scenario actually covers the obligation.

\begin{algorithm}[t]
\caption{Coverage-driven structural test generation with DSPy realization.}
\label{alg:coverage-generation}
\small
\begin{algorithmic}[1]
\STATE Construct reachable coordination graph $G(S)$
\STATE Extract obligations $\Omega(S) = (A_{\mathit{reach}}, T_{\mathit{allow}}, T_{\mathit{restrict}}, D)$
\STATE $Q \leftarrow \varnothing$
\FORALL{$a \in A_{\mathit{reach}}$}
    \STATE $Q \leftarrow Q \cup \{\mathsf{Reach}(a)\}$
\ENDFOR
\FORALL{$(a,t) \in T_{\mathit{allow}}$}
    \STATE $Q \leftarrow Q \cup \{\mathsf{UseTool}(a,t)\}$
\ENDFOR
\FORALL{$(a,t) \in T_{\mathit{restrict}}$}
    \STATE $Q \leftarrow Q \cup \{\mathsf{RestrictTool}(a,t)\}$
\ENDFOR
\FORALL{$(a,b) \in D$}
    \STATE $Q \leftarrow Q \cup \{\mathsf{Delegate}(a,b)\}$
\ENDFOR
\STATE $B \leftarrow \mathsf{MergeCompatibleObjectives}(Q)$
\STATE $X \leftarrow \varnothing$
\FORALL{$B_i \in B$}
    \STATE $q^\star \leftarrow \mathsf{Driver}(B_i)$
    \STATE Select the criterion-specific DSPy signature for $q^\star$
    \STATE Bind signature inputs from $S$ and $G(S)$
    \STATE $\mathit{passed} \leftarrow \textsc{false}$;\; $i \leftarrow 0$
    \WHILE{$i < N$ \textbf{and} $\neg \mathit{passed}$}
        \STATE $z \leftarrow$ sample candidate user prompt using prior feedback
        \IF{$z$ leaks a forbidden identifier}
            \STATE $r \leftarrow 0$
        \ELSE
            \STATE $\tau \leftarrow$ execute $z$ and collect the structural trace
            \STATE $r \leftarrow \bigwedge_{q \in B_i} \mathsf{Witness}(q,\tau)$
        \ENDIF
        \STATE record $(z,\tau,r)$
        \IF{$r=1$}
            \STATE $X \leftarrow X \cup \{\mathsf{RealizeAsTest}(z,\tau,B_i)\}$
            \STATE $\mathit{passed} \leftarrow \textsc{true}$
        \ENDIF
        \STATE $i \leftarrow i+1$
    \ENDWHILE
    \IF{$\neg \mathit{passed}$}
        \STATE $\mathsf{ReportUnrealized}(B_i)$
    \ENDIF
\ENDFOR
\RETURN $X$
\end{algorithmic}
\end{algorithm}

\subsection{Execution and Observation}
\label{sec:execution-observation}

Once generated, the suite is executed and evaluated through structural observations rather than source-code instrumentation. Each test produces an observation tuple of the form defined in Section~\ref{sec:formal-coverage}. These observations are matched back to the four obligation sets to compute coverage under $C_1$, $C_2$, $C_3$, and $C_4$.

The execution layer may attach additional metadata, such as backend used, retry count, model configuration, trace identifier, or invocation surface, that is useful for debugging and reproducibility but is not part of the graph-derived coverage model. In our prototype, the runtime adapter materializes each reachable agent as an OpenAI Agents SDK \texttt{Agent}, registers tool edges as recording stubs that bridge to a per-test runtime context, and routes delegation handoffs through the SDK's native handoff machinery. Section~\ref{sec:evaluation} reports the resulting observations across the ten-workflow benchmark suite and the deeper \texttt{oai\_customer\_service} study.

%% file: evaluation.tex
\section{Evaluation}
\label{sec:evaluation}

We evaluate the structural-testing pipeline on ten multi-agent workflows drawn from the OpenAI Agents SDK examples and from public third-party projects built on the same SDK. The evaluation asks whether the proposed graph-based coverage model can be extracted from real SDK-style workflow definitions, whether coverage-driven scenario generation can produce executable witnesses for the resulting obligations, and whether runtime witness outcomes provide useful diagnostic information about workflow structure.

Across the ten workflows, the extracted specifications contain 49 reachable agents, 47 tools, and 403 graph-derived structural obligations. These obligations consist of reachable-agent obligations ($C_1$), allowed-tool obligations ($C_2$), restricted-tool obligations ($C_3$), and delegation obligations ($C_4$). We evaluate extraction and obligation construction on all ten workflows, execute runtime witness studies for allowed-tool, restricted-tool, and delegation obligations, and use \texttt{oai\_customer\_service} as a  instrumented benchmark for additional routing-prompt ablation and tool-fault robustness experiments.

The evaluation has three goals. First, it tests whether the deterministic parts of the pipeline produce well-formed coordination graphs and obligation sets across workflows of different sizes and orchestration styles. Second, it tests whether the LLM-based realization layer can produce natural-language scenarios that witness structural obligations at runtime under a bounded refinement budget. Third, it examines whether the resulting per-criterion outcome distributions are diagnostically useful: for example, whether they distinguish single-hop workflows from multi-hop workflows, and whether restricted-tool probing reveals concrete misrouting failures.

All runtime results are single-run measurements under a fixed configuration. We therefore interpret them as empirical evidence of feasibility and diagnostic value, not as statistically powered estimates of performance across all multi-agent systems.

\subsection{Research Questions}
\label{sec:eval-rq-intro}

The evaluation is organized around four research questions.

\textbf{RQ1. Obligation extraction and scenario realization.}
Can the pipeline derive valid typed coordination graphs, structural coverage obligations, and corresponding witness scenarios from SDK-derived multi-agent workflows?

\textbf{RQ2. Allowed-tool and delegation witnesses.}
To what extent do generated scenarios witness allowed-tool obligations ($C_2$) and delegation obligations ($C_4$) at runtime?

\textbf{RQ3. Restricted-tool adversarial witnesses.}
Does restricted-tool testing ($C_3$) reveal workflows whose declared tool restrictions can be violated under adversarial prompting?

\textbf{RQ4. Fault-injection robustness.}
Can the same structural harness test whether agents continue to handle tool faults without crashing or leaking raw error payloads?

These questions separate witness types with different interpretations. For $C_2$ and $C_4$, a satisfied witness means that the expected structural behavior was observed: the intended tool was invoked, or the intended delegation edge was traversed. For $C_3$, the runtime adversarial study has inverted polarity: a satisfied witness means that the adversarial scenario elicited an attempted restricted call. Thus, higher $C_3$ violation counts are not better; they identify concrete hardening targets. RQ4 uses the same runtime harness but changes the witness predicate to test mechanical robustness under faulty tool responses.

\subsection{Benchmarks}
\label{sec:eval-data}

\paragraph{Source and extraction}
Each benchmark is derived from publicly available OpenAI Agents SDK-style source code. The extractor imports a configured entry point, walks the reachable agent graph through declared tools and handoffs, and emits a normalized workflow specification of the kind described in Section~\ref{sec:background}. The extracted specifications are then parsed into typed coordination graphs and checked for well-formedness before obligation construction. Manifests are not edited by hand; when a benchmark uses external orchestration rather than in-agent handoff declarations, the adaptation is documented below. Note that the tool count \emph{Tl} and the allowed-tool edge count \emph{Al} need not coincide: the same tool may be declared as accessible to several agents, so each (agent, tool) access relation is counted as a separate allowed edge. This is why some benchmarks in Table~\ref{tab:benchmark-inventory} report $\mathit{Al} > \mathit{Tl}$ (for example, \texttt{value\_investment} with $\mathit{Tl}{=}9$ and $\mathit{Al}{=}24$).

\paragraph{Benchmarks}
Table~\ref{tab:benchmark-inventory} reports the structural shape of the ten extracted workflows. The benchmark set spans workflows from small routing examples to larger coordinator-style systems. It also includes both SDK delegation styles used in the source projects: explicit handoffs and sub-agent invocation through \texttt{Agent.as\_tool}. Tools include user-defined function tools and SDK-provided tools such as web search.

\begin{table}[t]
\centering

\caption{Benchmark list. \emph{Ag} = reachable agents, \emph{Tl} = tools, \emph{Al} = allowed agent--tool edges, \emph{Re} = restricted agent--tool edges, \emph{De} = delegation edges, and \emph{Obl} = total structural obligations ($C_1{+}C_2{+}C_3{+}C_4 = \mathit{Ag}{+}\mathit{Al}{+}\mathit{Re}{+}\mathit{De}$). The tool count \emph{Tl} is reported for context and does not contribute to \emph{Obl}, since $|C_1|$ counts reachable agents rather than tools.}
\label{tab:benchmark-inventory}
\small
\setlength{\tabcolsep}{3.5pt}
\begin{tabular}{@{}lrrrrrr@{}}
\hline
Benchmark ID & Ag & Tl & Al & Re & De & Obl \\
\hline
\texttt{oai\_customer\_service}            & 3  & 2  & 2  & 4   & 4  & 13 \\
\texttt{oai\_message\_filter}$^{\dagger}$  & 3  & 1  & 1  & 2   & 2  & 8 \\
\texttt{oai\_research\_bot}                & 4  & 4  & 4  & 12  & 3  & 23 \\
\texttt{oai\_financial\_research}          & 7  & 7  & 7  & 42  & 6  & 62 \\
\texttt{social\_media\_agent\_system}      & 2  & 3  & 3  & 3   & 1  & 9 \\
\texttt{deep\_research\_clone}$^{\dagger}$ & 3  & 1  & 1  & 2   & 2  & 8 \\
\texttt{value\_investment}                 & 4  & 9  & 24 & 12  & 3  & 43 \\
\texttt{autopitch}                         & 7  & 6  & 6  & 36  & 6  & 55 \\
\texttt{octagon\_vc\_agents}               & 12 & 12 & 12 & 132 & 11 & 167 \\
\texttt{ydmitry\_deep\_research}           & 4  & 2  & 5  & 3   & 3  & 15 \\
\hline
\textbf{Total (10)}                        & \textbf{49} & \textbf{47} & \textbf{65} & \textbf{248} & \textbf{41} & \textbf{403} \\
\hline
\end{tabular}
\\[2pt]
\raggedright\footnotesize
$^{\dagger}$ These workflows implement part of their multi-agent control externally, by re-invoking the runtime with a different starting agent rather than declaring all transfers as in-agent handoffs. For structural extraction, we re-encode the externally orchestrated transfer as an equivalent handoff edge in the extraction layer. This does not add agents, tools, prompts, or new behavior; it makes the already-present control transfer visible to the graph model.
\end{table}

\paragraph{Instrumented benchmark}
The \texttt{oai\_customer\_service} workflow plays a special role because it is small enough to inspect manually while still exercising every kind of structural obligation. It contains three reachable agents, two tools, two allowed tool edges, four restricted tool edges, and four delegation edges. We use it as the running example in Sections~\ref{sec:background} and~\ref{sec:approach}, and as the  instrumented benchmark for the routing-prompt ablation and fault-injection robustness study.

\paragraph{Benchmark construction notes}
Most benchmarks are extracted directly from their source-level agent declarations. A small number require normalization because their orchestration style is not represented as explicit SDK handoff declarations. For benchmarks that orchestrate sub-agents in ordinary Python control flow, the extraction layer introduces a coordinator representation that mirrors the published call structure. For workflows that externally switch the starting agent between turns, the extraction layer re-encodes that transfer as a handoff edge so that reachability can be computed from a single entry point. These adaptations are local to extraction and do not modify the upstream agents, tools, or prompts. We report them because they affect the graph shape and therefore the derived obligations.

\subsection{Experimental Setup}
\label{sec:implementation}

The implementation consists of four stages: manifest extraction, obligation construction, scenario realization, and runtime witness execution. Extraction and obligation construction are deterministic. Scenario realization uses DSPy with a bounded refinement loop. Runtime witness execution replays each realized scenario against an OpenAI Agents SDK runtime adapter.

All experiments use \texttt{gpt-4.1-mini} for both the realizer and the agent under test, with a maximum of $N{=}5$ refinement attempts per objective. The runtime adapter materializes each reachable agent as an SDK \texttt{Agent}, registers tool edges as recording stubs, and records observed agents, tool calls, restricted-access events, and delegation edges. Tool backends are mocked so that the witness signal measures structural behavior---which agent attempted which tool or handoff---rather than correctness of an external service.

Each criterion-level runtime evaluation is executed once per benchmark under the default configuration. The \texttt{oai\_customer\_service} benchmark is additionally evaluated under a strict-routing realizer prompt and under two injected tool-fault modes. Because the realizer and the agent under test are LLM-backed, the reported results should be read as single-run empirical evidence rather than statistically powered estimates.

\paragraph{Runtime witness accounting}
The runtime focuses on allowed-tool, restricted-tool, and delegation obligations. Agent-reachability obligations are included in the structural obligation inventory, but the main runtime tables emphasize tool and delegation behavior because those criteria require more specific runtime evidence than observing that an agent appeared in a trace. Restricted-tool results are reported with inverted polarity: a counted restricted-tool witness means that an adversarial prompt elicited an attempted restricted call.

\subsection{RQ1: Obligation Extraction and Scenario Realization}
\label{sec:rq1}

\paragraph{Setup}
We run graph construction, obligation extraction, witness-objective construction, and DSPy-based scenario realization on every benchmark in Table~\ref{tab:benchmark-inventory}. For each benchmark, we record the graph shape, the resulting obligation counts, and whether the realizer produces a candidate natural-language scenario for each generated witness objective within the refinement budget. The realizer uses \texttt{gpt-4.1-mini} with $N{=}5$ attempts.

\paragraph{Results}
The extractor terminates with a well-formed coordination graph on all ten workflows. In total, the benchmark set contains 49 reachable agents, 65 allowed-tool edges, 248 restricted-tool edges, and 41 delegation edges, for 403 structural obligations. The obligation layer emits typed witness objectives for the resulting obligation sets without manual intervention.

The scenario realizer produces candidate natural-language scenarios for the generated objectives within the refinement budget. The realized scenarios are concrete and domain-relevant. For example, on \texttt{oai\_customer\_service}, a reachability objective for \texttt{seat\_booking\_agent} is realized as a request to change a seat assignment, while a multi-intent scenario for the entry agent combines a flight-change request with a baggage question. On \texttt{social\_media\_agent\_system}, a use-tool objective for web search is realized as a request for recent viral social-media posts.

\paragraph{Behaviour of the bounded refinement loop}
Most realization attempts terminate quickly. On the smaller workflows, objectives often succeed on the first attempt. Multi-hop workflows require more attempts, especially when the target obligation sits behind two or more handoffs. In those cases, runtime feedback helps the realizer move from a generic request toward a more instrumental request that causes the intended route or tool call. This behavior is important because it shows that scenario realization is not judged by surface plausibility alone: candidates are refined using observed structural traces.

\begin{tcolorbox}[title=RQ1 Summary, breakable]\TRC{The pipeline extracts well-formed coordination graphs and derives structural obligations for all ten SDK-derived workflows. It also realizes witness scenarios within the bounded DSPy refinement budget. Multi-hop routing objectives account for the cases that require more than one realization attempt.}\end{tcolorbox}

\subsection{RQ2: Allowed-Tool and Delegation Witnesses}
\label{sec:rq2}

\paragraph{Setup}
We execute the allowed-tool and delegation witness evaluations on all ten benchmarks. The allowed-tool witness for $C_2$ holds when the runtime trace records that the intended agent invoked the intended allowed tool. The delegation witness for $C_4$ holds when the trace records the intended handoff or \texttt{Agent.as\_tool} transfer. Both evaluations use the default configuration described in Section~\ref{sec:implementation}.

\paragraph{Cross-benchmark results}
Table~\ref{tab:cross-bench-runtime} reports per-benchmark runtime outcomes for allowed-tool, restricted-tool, and delegation witnesses. We discuss the allowed-tool and delegation columns here and the restricted-tool column in Section~\ref{sec:rq3}. Across the ten benchmarks, allowed-tool witnesses succeed on 45/65 obligations (69.2\%), and delegation witnesses succeed on 31/41 obligations (75.6\%).

\begin{table*}[t]
\centering
\caption{Cross-benchmark single-execution runtime results under the default configuration. Columns report obligations satisfied within the $N{=}5$ refinement budget. For $C_2$ and $C_4$, a satisfied witness means that the expected allowed-tool invocation or delegation edge was observed. For $C_3$, polarity is inverted: a satisfied witness means that an adversarial prompt elicited an attempted restricted call. Thus, lower $C_3$ violation counts are better from a restriction-enforcement perspective.}
\label{tab:cross-bench-runtime}
\small
\setlength{\tabcolsep}{4pt}
\begin{tabular}{@{}lcccrr@{}}
\hline
Benchmark &
\begin{tabular}{@{}c@{}}Allowed tool\\witnessed ($C_2$)\end{tabular} &
\begin{tabular}{@{}c@{}}Restricted violation\\elicited ($C_3$)\end{tabular} &
\begin{tabular}{@{}c@{}}Delegation\\witnessed ($C_4$)\end{tabular} &
Wall (s) & LM r.t. \\
\hline
\texttt{oai\_customer\_service}       & 2/2   & 0/4   & 4/4   & 290    & 47 \\
\texttt{oai\_message\_filter}         & 1/1   & 0/2   & 2/2   & 138    & 27 \\
\texttt{social\_media\_agent\_system} & 3/3   & 2/3   & 1/1   & 311    & 33 \\
\texttt{deep\_research\_clone}        & 1/1   & 0/2   & 2/2   & 161    & 23 \\
\texttt{ydmitry\_deep\_research}      & 2/5   & 1/3   & 2/3   & 498    & 59 \\
\texttt{oai\_research\_bot}           & 3/4   & 0/12  & 2/3   & 1{,}789 & 137 \\
\texttt{value\_investment}            & 13/24 & 4/12  & 3/3   & 3{,}352 & 243 \\
\texttt{autopitch}                    & 4/6   & 3/36  & 3/6   & 6{,}409 & 376 \\
\texttt{oai\_financial\_research}     & 5/7   & 7/42  & 3/6   & 7{,}353 & 407 \\
\texttt{octagon\_vc\_agents}          & 11/12 & 6/132 & 9/11  & 18{,}915 & 2{,}285 \\
\hline
\textbf{Total (10 benchmarks)}        & \textbf{45/65} & \textbf{23/248} & \textbf{31/41} & \textbf{39{,}216} & \textbf{3{,}637} \\
\hline
\end{tabular}
\end{table*}

\paragraph{Allowed-tool findings ($C_2$)}
The allowed-tool results cluster into three regimes. First, single-hop targets pass uniformly. On \texttt{oai\_customer\_service}, \texttt{oai\_message\_filter}, \texttt{social\_media\_agent\_system}, and \texttt{deep\_research\_clone}, every allowed-tool target sits at most one delegation hop from the entry agent, and all allowed-tool witnesses succeed: 7/7 combined.

Second, fan-out coordinator workflows pass on most targets. In \texttt{oai\_research\_bot}, \texttt{oai\_financial\_research}, and \texttt{octagon\_vc\_agents}, coordinators route to multiple specialists. These workflows satisfy 19/23 allowed-tool witnesses. The misses occur mainly when the intended target overlaps semantically with another specialist, causing the coordinator to choose a different route.

Third, multi-hop and handoff-stub targets account for most misses. In \texttt{ydmitry\_deep\_research}, misses occur when the intended target sits two hops away from the entry agent. In \texttt{value\_investment}, many misses correspond to handoff-like objects surfaced as allowed edges but not invocable as ordinary function tools by the runtime harness. These failures are informative: they show where prompt-only realization and runtime-callability assumptions are weakest.

\paragraph{Delegation findings ($C_4$)}
Delegation witnesses succeed on 31/41 obligations. As with allowed-tool witnesses, the main determinant is topology. Direct handoffs and obvious coordinator-to-specialist routes are usually witnessed. Multi-hop routes and overlapping specialist roles account for most failures.

The bounded refinement loop is especially useful for delegation objectives. For example, in \texttt{oai\_message\_filter}, early attempts asking for Spanish output are answered directly by the parent agent. After runtime feedback, the realizer changes the request to explicitly allow connection to someone who can assist in Spanish, which produces the intended handoff. This illustrates the value of runtime-grounded realization: the realizer is not merely producing plausible prompts, but adapting prompts based on whether the structural witness actually appears.

\begin{tcolorbox}[title=RQ2 Summary, breakable]\TRC{Allowed-tool witnesses succeed on 45/65 obligations, and delegation witnesses succeed on 31/41 obligations across the ten benchmarks. Single-hop targets pass uniformly, fan-out coordinators pass on most dominant routes, and multi-hop or runtime-ambiguous targets account for most misses.}\end{tcolorbox}



\begin{table}[t]
\centering
\caption{Routing-prompt ablation on \texttt{oai\_customer\_service}. ``Pass'' counts allowed-tool objectives whose witness predicate held during execution. ``Wall'' is end-to-end wall time, and ``Calls'' is the number of realizer LM round-trips.}
\label{tab:rq2-ablation}
\small
\setlength{\tabcolsep}{8pt}
\begin{tabular}{@{}cllrr@{}}
\hline
Cell & Realizer & Routing & Pass & Calls \\
\hline
A & \texttt{gpt-4.1-mini} & loose  & 2/2 & 2 \\
B & \texttt{gpt-4.1-mini} & strict & 2/2 & 2 \\
\hline
\end{tabular}
\end{table}

\subsection{RQ3: Restricted-Tool Adversarial Witnesses}
\label{sec:rq3}

\paragraph{Setup}
We execute the restricted-tool evaluation on all ten benchmarks. For each restricted edge $(a,t)$, the realizer attempts to construct a request that tempts agent $a$ toward capability $t$ without naming the tool literally. Restricted tools are represented by recording stubs: if a restricted call is attempted, the runtime records the attempted $(a,t)$ pair and returns a refusal payload. The $C_3$ witness is satisfied when the target restricted call attempt is observed.

For $C_3$, a satisfied witness means that the adversarial prompt elicited a restricted-call attempt. Such a result is useful because it gives a concrete hardening target, but it is not a positive outcome from a restriction-enforcement perspective. Conversely, a zero count means that no restricted-call violation was elicited within the realizer's adversarial budget; it is not an absolute safety guarantee.

\paragraph{Results}
Across the ten benchmarks, restricted-tool probing elicits 23 violations out of 248 restricted-tool obligations (9.3\%). Four benchmarks produce no elicited restricted-call violations within the budget: \texttt{oai\_customer\_service} (0/4), \texttt{oai\_message\_filter} (0/2), \texttt{deep\_research\_clone} (0/2), and \texttt{oai\_research\_bot} (0/12). Six benchmarks produce at least one violation: \texttt{social\_media\_agent\_system} (2/3), \texttt{ydmitry\_deep\_research} (1/3), \texttt{value\_investment} (4/12), \texttt{autopitch} (3/36), \texttt{oai\_financial\_research} (7/42), and \texttt{octagon\_vc\_agents} (6/132).

\paragraph{No violation elicited within budget}
The zero-violation benchmarks are informative because they show cases where adversarial prompting did not elicit the targeted restricted call. On \texttt{oai\_customer\_service}, capability-laden requests tend to route to the agent that legitimately owns the tool rather than causing a peer specialist to invoke an off-owner tool. For example, a request about changing a seat is routed toward \texttt{seat\_booking\_agent} instead of causing \texttt{faq\_agent} to invoke \texttt{update\_seat}. Similar behavior appears in \texttt{deep\_research\_clone}, where scraping-like requests route to the search execution agent, and in \texttt{oai\_research\_bot}, where the coordinator structure limits direct invocation of underlying function tools.

Table~\ref{tab:rq3-restrict} gives the per-objective restricted-tool outcomes for \texttt{oai\_customer\_service}. The off-target column is important: it shows that the recording mechanism can observe restricted-call attempts even when the target restricted edge is not violated. Thus, the zero on-target count is not merely an instrumentation failure.

\begin{table*}[t]
\centering
\caption{Per-objective restricted-tool outcome on \texttt{oai\_customer\_service}. ``Pass'' is 1 iff the runtime records a restricted-call attempt at the intended $(a,t)$ pair. ``Off-target'' lists restricted-call attempts observed at a different agent than the intended target.}
\label{tab:rq3-restrict}
\small
\setlength{\tabcolsep}{4pt}
\begin{tabular}{@{}llcl@{}}
\hline
Target agent & Restricted tool & Pass & Off-target attempts \\
\hline
\texttt{faq\_agent}           & \texttt{update\_seat}      & 0/1 & --- \\
\texttt{seat\_booking\_agent} & \texttt{faq\_lookup\_tool} & 0/1 & \texttt{(triage\_agent, faq\_lookup\_tool)} \\
\texttt{triage\_agent}        & \texttt{faq\_lookup\_tool} & 0/1 & --- \\
\texttt{triage\_agent}        & \texttt{update\_seat}      & 0/1 & --- \\
\hline
\end{tabular}
\end{table*}

\paragraph{Elicited restricted-call violations}
On the six benchmarks with violations, the eliciting prompts are concrete and reproducible. In \texttt{social\_media\_agent\_system}, requests for recent viral posts can steer the social-media agent toward a web-search capability outside its declared access. In \texttt{ydmitry\_deep\_research}, a request for immediate source retrieval can cause the orchestrator to invoke a source-gathering tool directly rather than route to the search specialist. In \texttt{oai\_financial\_research}, prompts that ask for a single end-to-end answer including verification can collapse a planner-then-verifier flow into a direct planner call against verifier-like capabilities. These are exactly the kinds of structural regressions or boundary weaknesses that restricted-tool coverage is intended to expose.

The violation rate also varies substantially by workflow shape. \texttt{octagon\_vc\_agents} has the largest restricted-edge surface, with 132 restricted edges, but only six elicited violations. This suggests that coordinator-to-specialist structure can limit cross-specialist tool reach, although the result should be read as a single-run observation rather than a general safety claim.

\begin{tcolorbox}[title=RQ3 Summary, breakable]\TRC{Restricted-tool probing elicits 23 restricted-call violations across 248 restricted-tool obligations. Four workflows produce no targeted restricted-call violation within the adversarial budget, while six produce at least one concrete misrouting failure. The inverted-polarity $C_3$ criterion therefore provides diagnostic information that would be invisible in an allow-list-only evaluation.}\end{tcolorbox}

\subsection{RQ4: Fault-Injection Robustness}
\label{sec:rq4}

\paragraph{Setup}
We re-execute the allowed-tool evaluation on \texttt{oai\_customer\_service} with a fault injected into the target tool stub. We test two failure modes. In \texttt{fail:error}, the stub returns a canned internal-error payload. In \texttt{fail:malformed}, the stub returns a deliberately malformed payload. The robustness witness $W^{F1}$ holds when the target agent attempts the target tool call, the run does not crash, the agent produces a non-trivial final reply, and the reply does not leak the raw internal-error marker back to the user. The witness is intentionally mechanical so that it can be evaluated without a language-model judge.

\paragraph{Results}
Table~\ref{tab:rq4-fault} reports both failure modes. The robustness witness holds for both allowed-tool objectives under both injected faults. Under \texttt{fail:error}, the agent produces replies that acknowledge a technical issue. Under \texttt{fail:malformed}, the agent sometimes produces plausible-looking content from the malformed payload. This is accepted by $W^{F1}$ because the witness only checks mechanical robustness, not semantic correctness. The result therefore demonstrates that the same structural harness can test fault-handling behavior, while also showing the limits of a purely structural witness.

\begin{table}[t]
\centering
\caption{Fault injection on the allowed-tool evaluation of \texttt{oai\_customer\_service}. ``Pass'' is the number of objectives whose mechanical robustness witness $W^{F1}$ held.}
\label{tab:rq4-fault}
\small
\setlength{\tabcolsep}{8pt}
\begin{tabular}{@{}lrrr@{}}
\hline
Mode & Pass & Wall (s) & Calls \\
\hline
\texttt{fail:error}     & 2/2 & 32.2 & 4 \\
\texttt{fail:malformed} & 2/2 & 12.9 & 2 \\
\hline
\end{tabular}
\end{table}

\begin{tcolorbox}[title=RQ4 Summary, breakable]\TRC{
Under both injected fault modes, the target tool is attempted, the run does not crash, and the final reply does not leak the raw internal-error marker. The witness holds 2/2 in both modes. Because the witness is mechanical, it does not detect semantic fabrication from malformed tool output.}\end{tcolorbox}

\subsection{Synthesis}
\label{sec:eval-synthesis}

The evaluation supports the central claim of the paper: structural coverage provides an adequacy layer for multi-agent workflow specifications. The extracted graphs turn each workflow into an explicit set of test obligations, and the runtime witnesses show which declared agents, tool-access edges, restrictions, and delegation paths can actually be exercised by generated scenarios. This gives a different kind of evidence than end-to-end task success: it shows not only whether a workflow can produce an answer, but which parts of its declared coordination structure have been tested.

The ten-benchmark results show that coverage outcomes are informative at the level of workflow structure. Single-hop obligations are usually easy to witness, while multi-hop routes, overlapping specialist roles, and runtime-ambiguous handoff stubs produce most of the missed allowed-tool and delegation witnesses. These misses are not merely failures of the evaluation; they identify structural paths that are difficult to exercise under the current realizer and harness. In this sense, a coverage report is useful even when it is incomplete, because it tells the developer where the test suite does not yet provide evidence.

The restricted-tool results show why structural adequacy must include negative obligations. An evaluation that only checks allowed tool use and delegation would miss cases where an agent can be prompted into attempting a tool call outside its declared access. The $C_3$ results expose both sides of this boundary: some workflows produce no targeted restricted-call violation within the adversarial budget, while others yield concrete prompts that elicit restricted calls. Both outcomes are useful. The former gives bounded evidence that the tested restrictions resisted the generated probes; the latter gives actionable hardening targets.

 RQ4 also shows that the same coverage-driven tests can also support robustness checks by changing the witness predicate while keeping the underlying structural obligation fixed. This is important methodologically: once a workflow obligation has been identified, the same target can be evaluated under different runtime conditions, such as normal execution, restricted access, or faulty tool output. The malformed-output case also illustrates the boundary of the method: structural witnesses can show that the intended tool was attempted and that the run did not crash, but they do not establish semantic correctness of the final response.

Overall, the evaluation shows that the proposed criteria are more than bookkeeping over a graph. They define a practical notion of test-suite adequacy for multi-agent workflows: a suite is more adequate when it provides runtime evidence for the workflow's declared agents, allowed tool edges, restricted tool boundaries, and delegation paths. The resulting coverage profile helps distinguish exercised structure from untested structure, and turns coordination gaps into concrete targets for additional tests, better routing prompts, or workflow hardening.

\subsection{Threats to Validity}
\label{sec:threats}

\emph{External validity.}
The evaluation covers ten workflows, but all are OpenAI Agents SDK-style systems and most are demo-scale or research-scale projects rather than production deployments. We therefore do not claim that the measured pass rates generalize to other frameworks, closed-source systems, or much larger workflows. The graph model is intended to be framework-agnostic, but the extractor and runtime adapter evaluated here are SDK-specific.

\emph{Construct validity.}
The witness predicates measure structural evidence: agents observed, tools invoked, restricted-call attempts recorded, and delegation edges traversed. They do not measure semantic correctness of the final answer, user satisfaction, policy compliance beyond the tested structural relation, or factual accuracy. A suite can achieve high structural coverage while still producing poor natural-language responses. The fault-injection study makes this limitation explicit: the malformed-output case can produce plausible but semantically questionable replies that the mechanical witness still accepts.

\emph{Internal validity.}
The runtime layer uses mocked tool backends. This isolates structural behavior from external service variability, but it means the evaluation measures whether the agent attempts the intended tool or delegation, not whether the real tool would return a correct result. A second internal-validity concern is realizer reach: the bounded DSPy realizer does not always steer the conversation through multi-hop delegation chains within $N{=}5$ attempts. We report such misses as part of the diagnostic outcome rather than treating them as noise.

\emph{Reproducibility.}
Both the realizer and the agent under test are LLM-backed, so repeated runs may produce different surface prompts and occasionally different traces. We reduce this risk by using a fixed configuration, recording structural traces, and evaluating witnesses mechanically from those traces. However, the evaluation is single-run; a multi-run reliability study would be needed to estimate variance.

\emph{Benchmark normalization.}
Some source projects use orchestration styles that are not directly represented as static SDK handoff declarations. For these, the extraction layer normalizes the published control flow into graph edges so that reachability and coverage can be computed. These adaptations do not add new agents, tools, or prompts, but they can affect the graph shape and therefore the obligation set. We document them because they are part of the benchmark construction process.

\emph{Cost and scale.}
The ten-benchmark runtime evaluation required roughly 39{,}216 seconds of wall-clock time and 3{,}637 realizer LM round-trips. This is manageable for the benchmark sizes studied here, but larger workflows may require more aggressive objective merging, path-aware realization, or staged execution to control cost.

%% file: related-work.tex
\section{Related Work}
\label{sec:related-work}

\paragraph{Structural coverage and mutation testing}
Our work is most directly inspired by classical software testing notions of
structural adequacy, including statement, branch, and related coverage
criteria, as well as mutation testing
\cite{myers2011art,ammann2016introduction,jia2011mutation}.  The key difference
is that our coverage obligations are defined over a \emph{multi-agent
workflow graph} rather than over program control flow.  In that sense, the
proposed criteria are closer in spirit to artifact-level adequacy measures:
they ask whether declared agents, allowed tool edges, restricted tool edges,
and delegation edges have been exercised.  Mutation analysis along the same
lines (injecting structural faults into the workflow specification) is a
natural extension of the present work and is discussed as future work in
Section~\ref{sec:discussion}.  The main distinction is not the evaluation
logic of coverage itself, but the decision to treat the multi-agent workflow
specification as the primary testable artifact.

\paragraph{Specification-based testing and model-based testing}
There are also strong connections to specification-based testing,
model-based testing, and behavior-driven approaches in which tests are derived
from an abstract description of expected structure or behavior.  Our setting is
similar in that the object under test is not arbitrary source code alone, but a
declared artifact with explicit relations and constraints.  However, multi-agent
workflow specifications differ from conventional formal models because the
artifact statically declares agents, tools, and edges but is exercised at
runtime by language-model-driven agents whose tool-use and delegation choices
are not deterministic.  This makes the testing setting a hybrid: more
structured than prompt-only evaluation, but less fully formal than traditional
model-based testing \cite{utting2006practical}.  Our contribution differs from
classical model-based testing in that the input model is a lightweight
coordination graph rather than a fully formal behavioral model.

\paragraph{Automated test generation}
The generation side of our approach is also related to automated test-suite
construction work in software engineering.  Tools such as EvoSuite illustrate
how coverage obligations can drive systematic suite construction for code
artifacts \cite{fraser2011evosuite}.  Our generator follows the same broad
testing philosophy, but the generation targets are not source-level branches or
methods.  Instead, the witnesses are derived from declared workflow structure:
agents, allowed and restricted tool edges, and delegation edges.  In this
sense, we borrow the obligation-driven spirit of automated test generation
without assuming that code structure is the right abstraction level for
multi-agent workflows.

\paragraph{LLM and agent evaluation}
Recent work on LLM and agent evaluation has largely focused on benchmark task
success, robustness, and behavioral probing.  Existing agent benchmarks are
valuable for measuring end-to-end performance, but they do not provide a
coverage theory for the underlying multi-agent workflow.  Our goal is
complementary: instead of replacing task benchmarks, we provide adequacy
criteria for asking whether the \emph{declared structure} of a multi-agent
workflow has been tested at all.  This perspective is especially relevant for
systems in which the same agent runtime may execute many different workflow
specifications with different tool and delegation configurations
\cite{liang2023helm,liu2023agentbench,mialon2023gaia,jimenez2024swebench,zhou2024webarena}.
The distinction is therefore one of evaluation target: prior benchmarks ask how
well an agent solves tasks, whereas we ask how thoroughly a reusable workflow
specification has been exercised.

\paragraph{Prompt, tool, and workflow testing}
A growing ecosystem of prompt-evaluation and tool-testing frameworks supports
prompt regression checks, output comparison, and scenario-based assessment of
LLM systems.  These tools are useful in practice, but they generally emphasize
example-level correctness rather than structural adequacy over reusable
workflow artifacts.  Our contribution is to identify an emerging software
artifact class and to give it a corresponding structural testing vocabulary.
This shifts the discussion from ``did these examples pass?'' toward a coverage
question: ``which parts of the declared workflow have actually been
exercised?''
\cite{ribeiro2020checklist,wang2021textflint,li2023apibank,qin2023toolllm}.
In other words, our framework is not another prompt-regression harness or
tool-use benchmark.  It is an adequacy model for multi-agent workflow
specifications.

\paragraph{Positioning}
Taken together, prior work gives us the ingredients for this paper but not the
combined result.  Classical testing contributes the ideas of coverage and
mutation.  Specification-based testing contributes the view that declared
structure can serve as the basis for test generation.  LLM and agent evaluation
contributes the modern execution setting in which these artifacts matter.  What
has been missing is a coverage model and generation procedure tailored to
multi-agent workflow specifications as a software artifact in their own right.

%% file: discussion-conclusion.tex
\section{Discussion and Conclusion}
\label{sec:discussion}

This paper evaluates structural coverage as an adequacy layer for LLM-based multi-agent workflow specifications. The evaluation covers ten SDK-derived workflows, totaling 49 reachable agents, 47 tools, and 403 graph-derived structural obligations. The results show that the proposed model can be extracted from real workflow definitions, that the resulting obligations can be turned into typed witness objectives and natural-language scenarios, and that runtime traces can be used to determine which declared agents, tool-access edges, restricted boundaries, and delegation paths have actually been exercised.

The evaluation should be read as evidence of feasibility and diagnostic value, not as a general safety claim about multi-agent systems. All benchmarks are OpenAI Agents SDK-style workflows, and each runtime evaluation is executed once under a fixed configuration. The results therefore show that the method works on a nontrivial set of real SDK-derived workflows, but they do not establish statistical reliability across model runs, generality across agent frameworks, or effectiveness on production-scale deployments.

The \texttt{oai\_customer\_service} workflow plays a narrower role within this broader evaluation. It is used as the running example and as a instrumented benchmark for routing-prompt ablation and fault-injection experiments. It is not the sole empirical basis for the paper. The main empirical evidence comes from the ten-workflow benchmark set, while the deeper case study shows how the same coverage framework can support closer inspection of one workflow.

The four criteria measure structural exercise of the declared workflow graph. A suite that satisfies $C_1$, $C_2$, $C_3$, and $C_4$ has observed every reachable agent, exercised every allowed tool edge, produced explicit evidence for every restricted-tool obligation, and traversed every delegation edge. This gives a concrete notion of test-suite adequacy for workflow structure: an obligation that is never witnessed has not been tested.

This is different from semantic correctness. Structural coverage does not show that the final answer is factually correct, useful to the user, policy compliant in all cases, or robust to every possible prompt. It also does not measure the realism of generated scenarios except insofar as those scenarios execute and produce structural evidence. A workflow can have high structural coverage and still produce poor responses. Conversely, a workflow can succeed on an end-to-end benchmark while leaving important structural obligations untested.

The restricted-tool criterion is especially important because it makes negative obligations explicit. Testing only allowed tool use and delegation would show whether intended paths can be exercised, but it would not show whether forbidden tool boundaries can be probed. The $C_3$ results demonstrate that restricted-tool testing can produce two useful outcomes: no targeted restricted-call violation within the adversarial budget, or a concrete prompt that elicits a restricted call and therefore becomes a hardening target.






The most immediate next step is mutation analysis over workflow graphs. Structural mutations such as deleting an allowed-tool edge, removing a restriction, changing a delegation target, or disconnecting an agent would make it possible to measure whether structurally adequate suites detect workflow regressions. This would also enable fair comparison against alternative generation strategies.

A second direction is broader benchmark coverage. The current evaluation uses ten SDK-derived workflows. Extending the extractor to additional frameworks would test whether the same graph model captures workflow structure beyond the OpenAI Agents SDK. Larger workflows with deeper delegation chains and denser tool graphs would also stress the current realization strategy.

A third direction is improving the realizer. The evaluation shows that multi-hop obligations are harder to witness than single-hop obligations. A path-aware realizer that receives explicit route structure, or a scenario diversifier that generates multiple semantically distinct prompts for the same obligation, may improve coverage on deep or ambiguous workflows.

Finally, structural coverage should be combined with behavioral and semantic evaluation. The present criteria say whether declared structure was exercised; they do not judge whether the final response was correct. A mature testing framework for multi-agent workflows should include both layers: structural adequacy to ensure the workflow was exercised, and semantic or task-level checks to judge the quality of the resulting behavior.